\documentclass[secnumarabic,12pt,tightenlines,eqsecnum,floats,aps,amsmath,amssymb,nofootinbib,superscriptaddress,showpacs,prd]{revtex4-1}

\usepackage{amsmath,amsthm,amssymb,amsfonts}
\usepackage{mathcomp}
\usepackage{bm}

\usepackage{hyperref}
\numberwithin{equation}{section}
\theoremstyle{plain}
\newtheorem{theorem}{Theorem}

\newtheorem{proposition}[theorem]{Proposition}

\theoremstyle{definition}

\theoremstyle{remark}

\usepackage{todonotes}

\begin{document}

\title{Quantization of scalar fields coupled to point-masses}

\author{J. Fernando \surname{Barbero G.}}
\email[]{fbarbero@iem.cfmac.csic.es}
\affiliation{Instituto de
	Estructura de la Materia, CSIC, Serrano 123, 28006 Madrid, Spain}
\affiliation{Grupo de Teor\'{\i}as de Campos y F\'{\i}sica Estad\'{\i}stica, Instituto Universitario Gregorio Mill\'an
	Barbany, Universidad Carlos III de Madrid, Unidad Asociada al IEM-CSIC.}

\author{Benito A. \surname{Ju\'arez-Aubry}}
\email[]{pmxbaju@nottingham.ac.uk}
\affiliation{School of Mathematical Sciences, University of Nottingham, Nottingham NG7 2RD, United Kingdom}

\author{Juan \surname{Margalef-Bentabol}}
\email[]{juan.margalef@uc3m.es}
\affiliation{Instituto de
	Estructura de la Materia, CSIC, Serrano 123, 28006 Madrid, Spain}
\affiliation{Instituto Gregorio
	Mill\'an, Grupo de Modelizaci\'on y Simulaci\'on Num\'erica,
	Universidad Carlos III de Madrid, Avda. de la Universidad 30, 28911
	Legan\'es, Spain}

\author{Eduardo J. \surname{S. Villase\~nor}}
\email[]{ejsanche@math.uc3m.es}
\affiliation{Instituto Gregorio
	Mill\'an, Grupo de Modelizaci\'on y Simulaci\'on Num\'erica,
	Universidad Carlos III de Madrid, Avda. de la Universidad 30, 28911
	Legan\'es, Spain}
\affiliation{Grupo de Teor\'{\i}as de Campos y F\'{\i}sica Estad\'{\i}stica, Instituto Universitario Gregorio Mill\'an Barbany, Universidad Carlos III de Madrid, Unidad Asociada al IEM-CSIC.}

\date{June 25, 2015}

\begin{abstract}
We study the Fock quantization of a compound classical system consisting of point masses and a scalar field. We consider the Hamiltonian formulation of the model by using the geometric constraint algorithm of Gotay, Nester and Hinds. By relying on this Hamiltonian description, we characterize in a precise way the real Hilbert space of classical solutions to the equations of motion and use it to rigorously construct the Fock space of the system. We finally discuss the structure of this space, in particular the impossibility of writing it in a natural way as a tensor product of Hilbert spaces associated with the point masses and the field, respectively.

\bigskip

\end{abstract}

\maketitle

%\clearpage
%%%%%%%%%%%%%%%%%%%%%%%%%%%%%%%%%%%%%%%%%%%%%%%%%%%%%%%%%%%%%%%%%%
%
% INTRODUCTION
%
\section{Introduction}\label{intro}

The main motivation of this work is to understand, in rigorous terms, the Hamiltonian formulation and subsequent Fock quantization of linear systems consisting of fields coupled to point masses (and, eventually, other low-dimensional objects). An important question that we will answer in the paper regards the possibility of writing the Hilbert space of such a compound system as a tensor product of Hilbert spaces associated with the point masses and the field. The point masses are introduced to model external devices that can be used both to excite the system and to act as detectors sensitive to the ``field quanta''. In this last sense they can be thought of as generalizations of the Unruh-DeWitt particle detectors and similar devices, used in the discussion of quantum field theories in curved space-times and accelerated frames \cite{Unruh:1976db, DeWitt, UnruhWald}. We wish to emphasize from the start that the point masses --that we introduce already at the classical level-- have nothing to do, in principle, with the field quanta despite the fact that the latter are usually interpreted as particles or quasiparticles.

The types of systems that we consider here are related --but not equal-- to field theories defined in bounded spatial regions and share some features with them, as will be explained in the following. There are several important field theoretic models that display interesting and non-trivial behaviors when defined in such regions (or more generally in the presence of spatial boundaries). Among them we would like to mention Chern-Simons \cite{Witten} and Maxwell-Chern-Simons models in 2+1 dimensions \cite{Schonfeld,DeserJakiwTempleton1,DeserJakiwTempleton2}, Yang-Mills theories in 3+1 dimensions \cite{Sniatycki}, and general relativistic models such as the isolated horizons used to study black hole entropy in loop quantum gravity \cite{Ashtekar:1997yu,Ashtekar:1999wa}. Part of our work is motivated by the comments appearing in \cite{Sniatycki1} regarding the use of boundaries with the same purpose as the classical point-particles that we introduce here.

The standard approach to derive the Hamiltonian formulation for field theories, especially when gauge symmetries are present, relies on the methods developed by Dirac \cite{Dirac}. These are straightforward to use in the case of mechanical systems with a finite number of degrees of freedom and --in simple circumstances-- can be adapted to field theories if one is willing to accept a certain lack of mathematical precision (regarding, for instance, the functional spaces describing the field degrees of freedom). In the presence of boundaries, however, the naive implementation of the Dirac algorithm is awkward and often leads to incomplete or plainly wrong results (see \cite{nos} and references therein). This is even more so for the model that we consider in the paper where we have both boundaries and particle-like objects associated with them. Some technical details regarding the difficulties of implementing the Dirac algorithm for field theories can be found in \cite{Gotaythesis}.

These difficulties --and other important ones-- can be avoided by the use of the geometric constraint algorithm developed by Gotay, Nester and Hinds \cite{GotayNesterHinds,Gotaythesis,Gotay:1978dv}. This method provides a rigorous, geometric and global way to obtain the Hamiltonian description of field theories and pays due attention to functional analytic issues. In particular it provides a completely detailed description of the spaces where the Hamiltonian dynamics takes place and can thus be used as the starting point for quantization. This is especially useful for linear theories for which Fock quantization can be rigorously defined starting from the complexification of an appropriately defined real Hilbert space of classical solutions (along the lines described, for example, in \cite{Wald} in the context of quantum field theory in curved spacetimes). The precise construction of the Fock space is important in order to discuss the eventual factorization of the Hilbert space $\mathcal{H}$  of the system in the form $\mathcal{H}=\mathcal{H}_{\mathrm{masses}}\otimes\mathcal{H}_{\mathrm{field}}$ that would account for a clean separation between \textit{quantum} point particle and field degrees of freedom.

In the present paper we will study a model consisting of a finite length elastic string attached, at the ends, to point particles subject to harmonic restoring forces (in addition to the ones exerted by the string). Although similar systems have been considered in the literature \cite{Yurke,Lewis} our approach will concentrate on several mathematical issues ---relevant from the physical point of view--- that have not been discussed elsewhere. Specifically we will introduce a natural way to deal with this model by using a particular class of measures and the Radon-Nikodym (RN) derivatives defined with their help.  This measure theoretic approach is suggested by Mardsen and Hughes in \cite[page 85]{elasticity book}.

The methods and ideas put forward here can be exported to more complicated situations. For instance, it should be straightforward to generalize them to deal with higher dimensional systems where, in addition to point particles, other low-dimensional objects could be coupled. From the perspective of the standard quantum field theory in curved spacetimes, we are extending the usual approach by replacing the all-important Laplace-Beltrami operator by an elliptic operator defined with the help of a measure that combines physical features of both the field and point masses of the system. We would like to remark here that, as free theories are essential building blocks for the perturbative quantization of non-linear models (for instance, they play a central role to define the Fock spaces used in their description), it is important to understand them well as a first step to consider their quantization in the presence of boundaries and/or lower-dimensional objects.

The layout of the paper is the following: After this introduction, section \ref{GNH} will be devoted to the classical description of the system under consideration. In particular, in \ref{classical_description} we present a short discussion of the resolution of the evolution equations by separation of variables. This motivates the introduction of non trivial measures that play a relevant role in section \ref{Alternative_Lagrangian}, where we provide an alternative Lagrangian formulation. In section \ref{section_Hamiltonian} we obtain the Hamiltonian description by implementing the GNH algorithm \cite{GotayNesterHinds,Gotaythesis,Gotay:1978dv}. We provide the precise description of the functional spaces relevant for the model. As it will be shown these are generalizations of Sobolev spaces that can be understood in a neat way with the help of a scalar product defined in terms of appropriate measures. In section \ref{Fock} we use the Hamiltonian description of the system to build a Fock space and quantize. We will pay particular attention to the characterization of this Fock space as a tensor product of Hilbert spaces associated with the field and the particles. We end the paper with our conclusions in section \ref{conclusions} and one appendix where we give a number of useful mathematical results.

{\color{black} The reader interested only in the physical results of the paper should skip sections \ref{Alternative_Lagrangian} and \ref{section_Hamiltonian}, where a number of mathematical details are provided, and go directly from section \ref{classical_description} to the quantum description in terms of physical modes discussed in section \ref{Fock}.}

\section{Classical description}\label{GNH}

\subsection{Lagrangian and field equations}

Let us consider a model consisting of an elastic string of finite length (in $1+1$ dimensions) coupled to two point masses located at the ends and attached to springs of zero rest length. Both the string and the masses are subject to restoring forces proportional to the deviations from their equilibrium configurations. For definiteness we will consider that the motion of the system is longitudinal, although this is not essential. From a logical perspective the equations of motion for such a system should be obtained by analyzing with due care the forces acting on the string and the masses, and using Newton's laws. In practice, however, it is more convenient to use an action written in terms of an easily interpretable Lagrangian and formally derive them by computing its first variations. Let us, then, start from the action
\begin{equation}
S(u)=\int_{t_1}^{t_2} L(u(t),\dot{u}(t))\, \mathrm{d}t
\label{action}
\end{equation}
where the Lagrangian, for smooth $Q$ and $V$, has the following form
\begin{equation}
L(Q,V)=\frac{\lambda}{2}\langle V,V\rangle-\frac{\epsilon}{2}\langle Q', Q'\rangle-\frac{m^2}{2}\langle Q,Q\rangle+\sum_{x\in\{0,\ell\}} \left(\frac{M_x}{2}V(x)^2-\frac{k_x}{2}Q(x)^2\right)\,.
\label{lagrangian}
\end{equation}
In the preceding expression $\langle \cdot ,\!\cdot \rangle$ denotes the usual scalar product in the Hilbert space $L^2(0,\ell)$ defined with the help of the Lebesgue measure $\mu_L$, $\ell$ is the length of the unstretched string, $\lambda$ its longitudinal mass density, $\epsilon$ its Young modulus, $m^2>0$ is the spring constant per unit length associated with the restoring force acting directly on the string\footnote{We follow the custom of calling this constant $m^2$ because the squared mass of the quantum excitations of the usual Klein-Gordon field is proportional to it. The case $m=0$ differs slightly from the one that we study but can be approached with the same methods.}, $M_0$ and $M_\ell$ are the masses of the point particles and $k_0\,,k_\ell$ the elastic constants of the springs attached to them. The field $Q(x)$ represents the deviation of the string point labelled by $x$ from its equilibrium position. Spatial derivatives are denoted by primes and time derivatives by dots.

As it can be seen the Lagrangian has terms of ``field'' and ``particle'' types involving spatial derivatives of first order, at most. It is convenient at this point to choose units of length, time and mass such that $\ell=\epsilon=\lambda=1$ (which in particular implies that the speed of sound is $c^2:=\epsilon/\lambda=1$). Notice that by doing this we exhaust all the freedom in the choice of units so we will have to keep $\hbar$ explicit when quantizing the model. To remind the reader of this choice we will rename the remaining constants in the Lagrangian as $\widetilde{\omega}^2:=m^2/\epsilon\,,\widetilde{\omega}_j^2:=k_j/M_j\,,\mu_j:=M_{j}/\epsilon$ so that \eqref{lagrangian} becomes
\begin{equation}
L(Q,V)=\frac{1}{2}\langle V,V\rangle-\frac{1}{2}\langle Q', Q'\rangle-\frac{\widetilde{\omega}^2}{2}\langle Q,Q\rangle+\sum_{j\in\{0,1\}} \frac{\mu_j}{2}\Big(V(j)^2-\widetilde{\omega}_j^2Q(j)^2\Big)\,.
\label{lagrangian2}
\end{equation}

{\color{black} It is interesting to mention at this point that the positions of the point particles are not classical independent degrees of freedom as they are given by continuity by the position of the ends of the string. This fact will have an analogue in the quantum description of the model (non-factorization of the Fock space, see section \ref{Fock}).}

The equations of motion derived from \eqref{lagrangian2} are:
\begin{align}
\ddot{u}(x,t)-u''(x,t)+\widetilde{\omega}^2 u(x,t)&=0\,,\quad x\in(0,1)\,,\label{eqmotion}\\
\mu_0 \ddot{u}(0,t)-u'(0,t)+\mu_0\widetilde{\omega}_0^2 u(0,t)&=0\,, \label{eqmotion2}\\
\mu_1 \ddot{u}(1,t)+u'(1,t)+\mu_1\widetilde{\omega}_1^2 u(1,t)&=0\,. \label{eqmotion3}
\end{align}
As can be readily seen the time evolution of the deformation of the string is governed by the 1+1 Klein-Gordon equation whereas the point masses at the boundary points move under the combined force exerted by the springs and the string (given by the spatial derivatives at the boundary). It is important to notice at this point that the preceding equations are not conventional in the sense that \eqref{eqmotion2} and \eqref{eqmotion3} are not \emph{standard} boundary conditions because they involve second order time derivatives. We will see in section \ref{classical_description} that this feature qualitatively changes the type of eigenvalue problem that has to be solved to identify the normal modes and characteristic frequencies. Actually it renders the present problem quite non-trivial because the relevant eigenvalue equations \textit{are not} of the standard Sturm-Liouville type and hence the classical theorems that ensure that the normal modes form a complete set cannot be applied. There is however a workaround to prove such result that consists in introducing a different measure space. We anticipate now the result of this approach:

\begin{proposition}\label{proposition equivalence equations}
The equations \eqref{eqmotion}-\eqref{eqmotion3} are contained in the equations:
\begin{align}
&\ddot{u}-(\Delta_\mu-\widetilde{\omega}^2) u=0  &&\hspace*{-3cm}x\in[0,1]\,, \mu\textrm{-a.e.}\label{eqbulk_u}\\
&(-1)^j\frac{\mathrm{d}u}{\mathrm{d}\mu}(j)-A(j)u(j)=0 &&\hspace*{-3cm}j\in\{0,1\}\,,\label{boundary_cond}
\end{align}
that consist of a 1+1 dimensional Klein-Gordon equation on the interval $[0,1]$ subject to Robin boundary conditions written in terms of the Radon-Nikodym derivative $\frac{d}{d\mu}$ with respect to the measure $\mu= \alpha_0 \delta_0 + \mu_L + \alpha_1 \delta_1$. The parameters $\alpha_j$  are related to the physical parameters of the problem by $ \alpha_j(1-\alpha_j \mu_j(\widetilde{\omega}^2_j-\widetilde{\omega}^2)  )^2=\mu_j$,
$A(j):=(\widetilde{\omega}^2_j-\widetilde{\omega}^2)\sqrt{\mu_j\alpha_j}$, and the Laplace-like operator
\[
\Delta_\mu:=(1+C)\dfrac{\mathrm{d}^2}{\mathrm{d}\mu^2}
\]
is defined in terms of the function $C(j):=A(j)\alpha_j$, $C(x):=0$ if $x\not= 0,1$.
\end{proposition}

In the previous proposition $\mu\textrm{-a.e.}$ stands for ``$\mu$-almost everywhere'', (i.e. the equality can fail to be true in a set of zero $\mu$-measure, at most). The measure $\mu$ is defined in the Appendix (see Equation \ref{Diracdelta}) and the domain of $\Delta_\mu$, which specifies the regularity conditions on the solutions to the equations of motion, is given in Appendix \ref{section_appendix_laplacian}. Instead of giving a direct proof of this result, we will apply separation of variables to \eqref{eqmotion}-\eqref{eqmotion3}. In this process some issues will arise, the most important one being that the standard Laplacian is not self adjoint. This will lead to the introduction of a new self-adjoint Laplace-like operator (in an appropriate functional space) in terms of which we obtain equations \eqref{eqbulk_u} and \eqref{boundary_cond}.

\subsection{Classical description of the model: solving the field equations}\label{classical_description}
In this section we consider the resolution of the equations of motion \eqref{eqmotion}-\eqref{eqmotion3} by using the method of separation of variables. By writing $u(x,t)=X(x)T(t)$ we get
\begin{eqnarray}
        &&\ddot{T}=(\lambda-\widetilde{\omega}^2)T\label{ee2}\\
        &&X''=\lambda\,X\label{ee1}\\
        &&X'(0)=\mu_0(\lambda+\widetilde{\omega}_0^2-\widetilde{\omega}^2)X(0)\label{ee3}\\
        &&X'(1)=-\mu_1(\lambda+\widetilde{\omega}_1^2-\widetilde{\omega}^2)X(1)\label{ee4}
\end{eqnarray}
\noindent where $\lambda\in\mathbb{R}$. In this form these equations do define an eigenvalue problem for $X$ with one key (and relatively unusual) feature: the eigenvalue appears also in the boundary equations \eqref{ee3}, \eqref{ee4}. This means that we are not directly dealing with a Sturm-Liouville problem and, hence, we cannot directly import the usual results that characterize the eigenvalues $\lambda$ (do they exist? are they isolated? are they bounded?) and the corresponding eigenfunctions (are they a complete set? are they orthogonal?). The answer to these questions is important in order to expand the general solution to the equations of motion as a functional series of eigenfunctions and also to quantize the system.

In any case, a lot of information can be gathered in practice by solving the concrete eigenvalue problem that we have at hand so we sketch now the computation of the eigenvalues and the eigenfunctions.

\bigskip

\noindent \textit{i) Negative eigenvalues} $\lambda=-\omega^2<0$\vspace*{1ex}

\noindent The solutions to \eqref{ee1} are of the form $X_\lambda=a\cos(\omega x)+b\sin(\omega x)$ with $a,b\in\mathbb{R}$. The conditions \eqref{ee3},\eqref{ee4} imply that
\begin{eqnarray*}
& &\mu_0(\omega^2-\Delta\widetilde{\omega}^2_0)a+\omega b=0\\
& & \left(\rule{0ex}{2.5ex}\mu_1(\omega^2-\Delta\widetilde{\omega}^2_1)\cos\omega+\omega\sin\omega\rule{0ex}{2.5ex}\right) a+\left(\rule{0ex}{2.5ex}\mu_1(\omega^2-\Delta\widetilde{\omega}^2_1)\sin\omega-\omega\cos\omega\rule{0ex}{2.5ex}\right)b=0
\end{eqnarray*}

Where we have introduced the shorthand $\Delta\widetilde{\omega}^2_j:=\widetilde{\omega}^2_j-\widetilde{\omega}^2$. These have non-trivial solutions for $a$, $b$ if and only if
\begin{equation}
\hspace*{-2ex}\left(\rule{0ex}{2.5ex}\omega^2\!\!-\!\mu_0\mu_1(\omega^2\!\!-\!\Delta\widetilde{\omega}^2_0)(\omega^2\!\!-\!\Delta\widetilde{\omega}^2_1)\right)\sin\omega+\left( \mu_0(\omega^2\!\!-\!\Delta\widetilde{\omega}^2_0)+\mu_1(\omega^2\!\!-\!\Delta\widetilde{\omega}^2_1)\rule{0ex}{2.5ex}\right)\omega\cos\omega=0\,.\label{eq_neg_eigen}
\end{equation}
It is straightforward to see that \eqref{eq_neg_eigen} has an infinite number of solutions for $\omega$ for every (physical) choice of parameters $\mu_0\,,\mu_1\,,\widetilde{\omega}_0^2\,,\widetilde{\omega}_1^2\,,\widetilde{\omega}^2$; in fact, there exists $n_0\in\mathbb{N}$, such that every interval of the form $(k\pi,(k+1)\pi)$ contains one and only one solution of \eqref{eq_neg_eigen} for every $k\in\mathbb{N}$, $k>n_0$. In the asymptotic limit $k\rightarrow\infty$ we have
\begin{equation}
\omega_k=k\pi+\left(\rule{0ex}{2.5ex} \frac{1}{\mu_0}+\frac{1}{\mu_1}\rule{0ex}{2.5ex}\right)\frac{1}{k\pi}+O\big(k^{-2}\big)\,,
\label{asympt}
\end{equation}
so we see that we actually have an infinite set of negative eigenvalues and also that they grow without bound. Finally the eigenfunctions have the form (labelling them with $k$ and with a minus superscript to indicate that the eigenvalue is negative)
\[
X_k^-(x)=\omega_k \cos(\omega_k x)+\mu_0(\Delta\widetilde{\omega}^2_0-\omega_k^2)\sin(\omega_kx)\,.
\]

\bigskip

\noindent \textit{ii) Positive eigenvalues} $\lambda=\omega^2>0$\vspace*{1ex}

\noindent In this case the eigenfunctions $X_\lambda(x)$ have the form $X_\lambda(x)=ae^{\omega x}+be^{-\omega x}$ where the real coefficients $a$, $b$ must satisfy now the conditions
\begin{eqnarray*}
&& \left(\rule{0ex}{2.5ex}\omega-\mu_0(\omega^2+\Delta\widetilde{\omega}^2_0)\rule{0ex}{2.5ex}\right)a-\left(\rule{0ex}{2.5ex}\omega
+\mu_0(\omega^2+\Delta\widetilde{\omega}^2_0)\rule{0ex}{2.5ex}\right)b=0\\
& &  e^\omega\left(\rule{0ex}{2.5ex}\omega+\mu_1(\omega^2+\Delta\widetilde{\omega}^2_1)\rule{0ex}{2.5ex}\right)a-e^{-\omega}\left(\rule{0ex}{2.5ex}\omega-\mu_1(\omega^2+\Delta\widetilde{\omega}^2_1)\rule{0ex}{2.5ex}\right)b=0
\end{eqnarray*}
that have non-trivial solutions if and only if
\begin{eqnarray*}
&&e^{-\omega}\left(\rule{0ex}{2.5ex}  \omega-\mu_0(\omega^2+\Delta\widetilde{\omega}^2_0)\right)\left(\rule{0ex}{2.5ex} \omega-\mu_1(\omega^2+\Delta\widetilde{\omega}^2_1)\right)\hspace*{6cm}\label{eq_pos_eigen}\\
&&\hspace*{3.2cm}-e^\omega\left(\rule{0ex}{2.5ex}  \omega+\mu_0(\omega^2+\Delta\widetilde{\omega}^2_0)\right)\left(\rule{0ex}{2.5ex} \omega+\mu_1(\omega^2+\Delta\widetilde{\omega}^2_1)\right)=0\,.
\end{eqnarray*}
It can be seen that this equation has a finite number of solutions $N$ (maybe none) depending on the particular choices of the physical parameters defining the problem (e.g. if both $\Delta\widetilde{\omega}^2_j\geq0$, there is no positive eigenvalue). Notice that, as the energy is constant, the function $T$ cannot have an exponential growth so $a\leq\widetilde{\omega}^2$ and, in fact, the limit case $a=\widetilde{\omega}^2$ (where $T$ has a linear behavior) happens only if $\widetilde{\omega}_j=0=\widetilde{\omega}$ (in the paper we are assuming $\widetilde{\omega}>0$). Consider now $k\in\{-1,\ldots,-N\}$ as a negative and finite counter, the corresponding eigenfunctions are
\[
X^+_k(x)=\big(\omega_k+\mu_0(\omega_{k}^2+\Delta\widetilde{\omega}^2_0)\big)e^{\omega_k x}+\big(\omega_k-\mu_0(\omega_k^2+\Delta\widetilde{\omega}^2_0)\big)e^{-\omega_k x}\,.
\]

\bigskip

\noindent\textit{iii) Zero mode }$\lambda=0$\vspace*{1ex}

\noindent It is easy to check that $\lambda=0$ is an eigenvalue if and only if $(1+\mu_0\Delta\widetilde{\omega}^2_1)(1+\mu_1\Delta\widetilde{\omega}^2_1)=1$, in which case the solution is simply $X^0(x)=X^0(0)(1+\mu_0\Delta\widetilde{\omega}^2_0x)$. This can only happen if $\Delta\widetilde{\omega}^2_0=0=\Delta\widetilde{\omega}^2_1$ or $\Delta\widetilde{\omega}^2_0<0<\Delta\widetilde{\omega}^2_1$ or $\Delta\widetilde{\omega}^2_1<0<\Delta\widetilde{\omega}^2_0$.

\bigskip

From now on we will collectively denote the eigenvectors as $X_n$ with $X_n=X^+_n$ when $n\in\{-1,\ldots,-N\}$, $X_0=X^0$ (``when it exists''), and  $X_n=X^-_n$ when $n\in \mathbb{N}$. Their associated eigenvalues $\lambda_n$ are, respectively $\lambda_n=\omega^2_n$ when $n\in\{-1,\ldots,-N\}$, $\lambda_0=0$, and $\lambda_n=-\omega^2_n$ when $n\in \mathbb{N}$.

Notice that, at this point, we have found the solutions of the form $u_n(x,t)=X_n(x)T_n(t)$ to equations \eqref{ee2}-\eqref{ee4}. In order to find all the solutions to \eqref{eqmotion}-\eqref{eqmotion3} we have to prove that the eigenfunctions $\{X_k\}$ form a \emph{complete set} in an appropriate functional space. This is not straightforward because, as mentioned before, we do not have a standard Sturm-Liouville problem.

\bigskip

The generalized Sturm-Liouville problems of the form defined by \eqref{ee1}-\eqref{ee4} have a long history both in mechanics and mathematics (see, for instance \cite{Yurke,Walecka,Churchill,Evans} and references therein). A direct but important observation is the fact that eigenfunctions $X_k$, corresponding to different eigenvalues $\lambda_k$, \emph{are not orthogonal} with respect to the standard scalar product in $L^2(0,1)$ but are orthogonal with respect to the following modified scalar product \cite{Churchill}
\begin{equation}
\langle\!\langle u,\,v \rangle\!\rangle:=\mu_0u(0)v(0)+\mu_1u(1)v(1)+\int_0^1 u\cdot v\, \mu_L\,.
\label{scalar_product}
\end{equation}
This can be readily proved by taking two such eigenfunctions $X_m$, $X_n$ (associated with different eigenvalues $\lambda_m$, $\lambda_n$), integrating the following identity over the interval $[0,1]$
\[
(X_n X'_m-X_m X'_n)'=(\lambda_m-\lambda_n)X_m X_n\,
\]
and using the boundary conditions. The appearance of the scalar product \eqref{scalar_product} suggests to look for Hilbert spaces adapted to it. We consider this issue in the next section.

\bigskip

{\color{black} We want to point out here that, at this stage, the position of the point particles are the limits $X(j)=\lim_{x\to j} X(x)$. However, as will be shown in the following, there are technical reasons to introduce function spaces where generically this equality does not hold (instead, boundary conditions of the type \eqref{eq robin-like equivalent} are satisfied).}

\subsection{Alternative Lagrangian for the system}\label{Alternative_Lagrangian}

In order to get a precise Hamiltonian formulation for the dynamics of the system at hand, properly identify and characterize its degrees of freedom and deal with the delicate analytic and geometric issues posed by the presence of boundaries, it is most appropriate to use the GNH geometric algorithm developed in \cite{GotayNesterHinds,Gotaythesis,Gotay:1978dv}. A convenient starting point is to introduce a new Lagrangian defined on a \textit{manifold domain} \cite{ChernoffMarsden} of the tangent bundle of a configuration space, that we will take to be a real Hilbert space. In practice this means that we will have to extend the system somehow and also consider field configurations which are less smooth than the ones used in \eqref{lagrangian2}. We will require, nonetheless, that the solutions to  \eqref{eqmotion}-\eqref{eqmotion3} are appropriately contained in those corresponding to the equations of motion of the new Lagrangian.

In view of the results of previous section, instead of working with \eqref{lagrangian2}, it is natural to look for a generalized Lagrangian written in terms of the scalar product defined by a certain measure $\mu$ and the associated RN derivatives. The hope --that will be realized-- is that the equations of motion give rise to a
\textit{standard Sturm-Liouville problem} in terms of this derivative and also that the boundary conditions defining the elliptic operator that will play a central role in its solution are such that its self-adjointness (and other related properties such as the completeness of the set of eigenfunctions) can be readily asserted and proved.

Let us consider then
\[
\mathcal{D}:=\left\{u\in L^2_\mu[0,1]:\,\exists\dfrac{\mathrm{d}u}{\mathrm{d}\mu}(x) \ \forall x\in[0,1]\,, \dfrac{\mathrm{d}u}{\mathrm{d}\mu}{\mathrm{\ is}\,\mu\textrm{-a.c.}}\,, \dfrac{\mathrm{d}^2u}{\mathrm{d}\mu^2}\in L^2_\mu[0,1]\right\}
\]
 and the Lagrangian $L:\mathcal{D}\times L^2_\mu[0,1]\rightarrow \mathbb{R}$ of the following form
\begin{eqnarray}
L(Q,V)&=&\frac{1}{2}\langle V,V\rangle_\mu-\frac{1}{2}\left\langle \frac{\mathrm{d}Q}{\mathrm{d}\mu},\frac{\mathrm{d}Q}{\mathrm{d}\mu}\right\rangle_{\hspace*{-1ex}\mu\hspace*{0.1ex}}-\frac{1}{2}\widetilde{\omega}^2\langle Q,Q\rangle_\mu\label{new_lagrangian}\\
&-&\sum_{j\in\{0,1\}}\alpha_j^2\left((-1)^j\frac{\mathrm{d}Q}{\mathrm{d}\mu}(j)-A(j)Q(j)\right)\frac{\mathrm{d}^2Q}{\mathrm{d}\mu^2}(j)-\frac{1}{2}\sum_{j\in\{0,1\}}A(j)Q(j)^2\,,\nonumber
\end{eqnarray}
where  $\langle \cdot ,\!\cdot \rangle_\mu$ is the scalar product with respect to the measure $\mu=\alpha_0\delta_0+\left.\mu_L\right|_{(0,1)}+\alpha_1\delta_1$, $\frac{\mathrm{d}}{\mathrm{d}\mu}$ denotes the associated RN derivative and $\mu\textrm{-a.c.}$ stands for ``$\mu$-absolutely continuous''. In view of the scalar product \eqref{scalar_product}, one might naively expect that $\alpha_j=\mu_j$, {\color{black} however $\alpha_j$, as well as $A(j)$, have to be taken as} non-trivial functions of the physical parameters of the model (see \cite{Evans} and appendix \ref{Hilbert_spaces} for more details, in particular equations \eqref{eq robin-like equivalent} and \eqref{eq_equivalence_scalar}).

The equations of motion can be obtained by computing the first variation of the action derived from the Lagrangian. A straightforward computation gives
\begin{eqnarray*}
&&\delta S(u)=\int_{t_1}^{t_2}\mathrm{d}t \Bigg(-\langle  \ddot{u},\delta u\rangle_\mu+\langle\Delta_\mu u,\delta u\rangle_\mu-\widetilde{\omega}^2\langle u,\delta u\rangle_\mu\\
&&\hspace*{4cm}-\sum_{j\in\{0,1\}}\Big((-1)^j\alpha_j\gamma_j(\delta u')-\gamma_j(\delta u)\Big)\Big((-1)^j\frac{\mathrm{d}u}{\mathrm{d}\mu}(j)-A(j)u(j)\Big)\Bigg)\,,
\end{eqnarray*}
where $\delta u\in\mathcal{D}$ so that the traces $\gamma_j(\delta u')$ and $\gamma_j(\delta u)$ are well defined (see Appendix \ref{appendixB2}). From this last expression we get the equations of motion
\begin{align}
&\ddot{u}-(\Delta_\mu-\widetilde{\omega}^2) u=0  &&\hspace*{-3cm}x\in[0,1]\,, \mu\textrm{-a.e.}\label{eqbulk_u_2}\\
&(-1)^j\frac{\mathrm{d}u}{\mathrm{d}\mu}(j)-A(j)u(j)=0 &&\hspace*{-3cm}j\in\{0,1\}\,.\label{boundary_cond_2}
\end{align}
They have the form of the 1+1 dimensional Klein-Gordon equation on the interval $[0,1]$ subject to Robin boundary conditions written in terms of the RN derivative. It is important to remember that $\Delta_\mu$ \emph{is not self-adjoint} in $\mathcal{D}$. However, the solutions to \eqref{boundary_cond_2} belong to
\[
\widehat{\mathcal{D}}:=\left\{u\in \mathcal{D}:(-1)^j\dfrac{\mathrm{d}u}{\mathrm{d}\mu}(j)-A(j)u(j)=0\right\}
\]
and in this domain $\Delta_\mu$ is indeed self adjoint (see section \ref{section_appendix_laplacian}).

In order to see that these equations describe the same dynamics as \eqref{eqmotion}-\eqref{eqmotion3} we first notice that in the open interval $(0,1)$ equation \eqref{eqbulk_u_2} is simply the Klein-Gordon equation $\ddot{u}-u''+\widetilde{\omega}^2u=0$. If we write now the boundary conditions \eqref{boundary_cond_2} in the form given by \eqref{eq robin-like equivalent}, plug the resulting expression into $\ddot{u}(j)-\Delta_\mu u(j)+\widetilde{\omega}^2 u(j)=0$, and use the relations that fix $\alpha_j$ and $A(j)$ in terms of the physical parameters for the problem we immediately obtain \eqref{eqmotion2}-\eqref{eqmotion3}. This completes the proof of Proposition \ref{proposition equivalence equations}.

\subsection{Hamiltonian formulation}\label{section_Hamiltonian}

The equations \eqref{eqbulk_u_2} and \eqref{boundary_cond_2} derived from the Lagrangian \eqref{new_lagrangian} can be understood as a particular case of the abstract wave equation (see discussion in \cite{nos}). By using the results described there, it is possible to directly get both the expression of the Hamiltonian vector field and the manifold domains where the dynamics takes place.

\begin{proposition} The Hamiltonian dynamics  takes place in the second class\footnote{A submanifold $N\stackrel{\jmath}{\rightarrow} M$ of a presymplectic manifold $(M,\omega)$ is said to be \textit{second class} if $TN^\perp\cap\underline{TN}=\{0\}$ where $TN^\perp:=\{Z\in TM|_N:\omega|_N(Z,X)=0\,\forall X\in\underline{TN}\}$, and $\underline{TN}:=j_*TN$.} (generalized) submanifold $\mathcal{M}_2=\widehat{\mathcal{D}}\times\mathcal{D}$  of the weakly symplectic manifold $(\mathcal{M}_1=\mathcal{D}\times L^2_\mu,\omega)$, where $\omega$ is the  pullback to $\mathcal{M}_1$ of the strong, canonical, symplectic form on $L^2_\mu\times L^2_\mu$. The dynamics is governed by the Hamiltonian vector field $X=(X_Q,X_P): \mathcal{M}_2\rightarrow \mathcal{M}_1$ whose components are
\begin{eqnarray}
X_Q(Q,P)&=&P\\
X_P(Q,P)&=&-(\widetilde{\omega}^2-\Delta_\mu)Q\,,
\end{eqnarray}
and the Hamiltonian is given by
\begin{eqnarray}
H(Q,P)&=&\frac{1}{2}\langle P,P\rangle_\mu+\frac{1}{2}\left\langle \frac{\mathrm{d}Q}{\mathrm{d}\mu},\frac{\mathrm{d}Q}{\mathrm{d}\mu}\right\rangle_{\hspace*{-1ex}\mu\hspace*{0.1ex}}+\frac{1}{2}\widetilde{\omega}^2\langle Q,Q\rangle_\mu\label{Hamiltonian}\\
&+&\sum_{j\in\{0,1\}}\alpha_j^2\Big((-1)^j\frac{\mathrm{d}Q}{\mathrm{d}\mu}(j)-A(j)Q(j)\Big)\frac{\mathrm{d}^2Q}{\mathrm{d}\mu^2}+\frac{1}{2}\sum_{j\in\{0,1\}}A(j)Q^2(j)\,.\nonumber
\end{eqnarray}
\end{proposition}

As we have mentioned before, this result follows directly from the abstract wave equation \cite{nos}, however, there are interesting technical details in the derivation of the Hamiltonian formulation by using the GNH algorithm. For completeness, we sketch here how this works for our system.

We start by computing the fiber derivative defined by the Lagrangian \eqref{new_lagrangian}
\begin{equation}
FL:\mathcal{D}\times L^2_\mu\rightarrow   L^2_\mu \times L^{2*}_\mu:(Q,V)\mapsto (Q,\langle V,\cdot\rangle_\mu)\,.
\label{fiber-derivative}
\end{equation}
In order to conform with the standard notation we will write it as $(Q,\langle P,\cdot\rangle_\mu)$ with $P:=V$. By using now the Riesz representation theorem we can simply consider, as in the standard case of the scalar field with Dirichlet or Robin boundary conditions \cite{nos}, that the fiber derivative is $\mathcal{D}\times L^2_\mu\rightarrow   L^2_\mu \times L^2_\mu$ with $FL(Q,V)=(Q,V)$ and the primary constraint manifold is $\mathcal{M}_1:=\mathcal{D}\times L^2_\mu$ taken as a generalized submanifold of $L^2_\mu\times L^2_\mu$.

The space $L^2_\mu\times L^2_\mu$ carries a canonical strongly nondegenerate symplectic form (inherited from the cotangent bundle $L^2_\mu\times L^{2*}_\mu$) given by
\begin{equation}
\Omega_{(Q,P)}((q_1,p_1),(q_2,p_2))=\langle q_1,p_2\rangle_\mu-\langle q_2,p_1\rangle_\mu
\label{symplectic_1}
\end{equation}
where we have $Q,P,q_i,p_i\in L^2_\mu$. The pull-back of $\Omega$ to $\mathcal{M}_1$ is the weakly symplectic form $\omega:=FL^*\Omega$ given by
\begin{equation}
\omega_{(Q,P)}((q_1,p_1),(q_2,p_2))=\langle q_1,p_2\rangle_\mu-\langle q_2,p_1\rangle_\mu\,,
\label{omega}
\end{equation}
where we have now $Q,q_i\in \mathcal{D}$ and  $P,p_i\in L^2_\mu$. It is interesting to mention that the ``boundary terms'' of the scalar product give rise here to ``boundary terms'' in the symplectic form. However we will see that this does not imply the existence of boundary degrees of freedom.

In order to obtain the Hamiltonian on $\mathcal{M}_1$ we compute the energy
\begin{equation}
H\circ FL(Q,V)=\langle V, V \rangle_\mu- L(Q,V)\,.
\label{HcircFL}
\end{equation}
This expression fixes the values of the Hamiltonian $H$ only on the primary constraint submanifold $\mathcal{M}_1$, however, this is the only information that we need to proceed with the GNH algorithm. From \eqref{HcircFL} we find that $H:\mathcal{M}_1\rightarrow \mathbb{R}$ is given by Equation \ref{Hamiltonian}.

On the primary constraint submanifold $\mathcal{M}_1$, vector fields are maps $X:\mathcal{M}_1\rightarrow \mathcal{M}_1\times \mathcal{M}_1:(Q,P)\mapsto((Q,P),(X_Q(Q,P),X_P(Q,P))$, such that $X_Q(Q,P)\in\mathcal{D}$ and $X_P(Q,P)\in L^2_\mu$. We have then
\[(i_X\omega)_{(Q,P)}(q,p)=\langle X_Q,p \rangle_\mu-\langle q,X_P \rangle_\mu\,.\]

We must find now a submanifold $\mathcal{M}_2$ and an injective immersion $\mathcal{M}_2\stackrel{\jmath_2}{\rightarrow}\mathcal{M}_1$ that allows us to solve the equation
\begin{equation}
(i_X\omega-\mathrm{d}H)|_{\jmath_2(\mathcal{M}_2)}=0\,.
\label{eqHam}
\end{equation}
Notice that this will require us to identify $\mathcal{M}_2$ as a subspace of $\mathcal{M}_1$ and also to specify its topology (by giving, for instance, a scalar product on it).

The resolution of Equation \eqref{eqHam} is relatively long but straightforward (see \cite{nos} for several similar computations). The final result is that $\mathcal{M}_2=\widehat{\mathcal{D}}\times\mathcal{D}$ and the Hamiltonian vector field is given by
\begin{eqnarray}
X_Q(Q,P)&=&P\\
X_P(Q,P)&=&-(\widetilde{\omega}^2-\Delta_\mu)Q\,.
\end{eqnarray}
The injection $\jmath_2:\mathcal{M}_2\rightarrow \mathcal{M}_1$ is just given by the inclusion map and is continuous if we use the natural topologies defined by the scalar products in $\mathcal{M}_1$ and $\mathcal{M}_2$. The Hamiltonian vector field $(X_Q,X_P)$ is also continuous in the same topology.

By using the same kind of argument employed in \cite{nos}, it is straightforward to show that the closure of $\widehat{\mathcal{D}}\times \mathcal{D}$ in $\widehat{\mathcal{D}}\times L^2_\mu$ satisfies $\mathrm{cl}_{\mathcal{D}\times L^2_\mu} (\widehat{\mathcal{D}}\times \mathcal{D})=\widehat{\mathcal{D}}\times L^2_\mu$. We then conclude that the Hamiltonian vector field $X$ is tangent to $\overline{\mathcal{M}_2}$ so the GNH algorithm stops. We finish this section with a comment. Although it is not obvious at first sight, the Hamiltonian \eqref{Hamiltonian} is positive definite when restricted to $\mathcal{M}_2$.

\section{Fock quantization}\label{Fock}

{\color{black} The fact that the positions of the point particles attached at the ends of the string are not independent physical degrees of freedom actually suggests that the Fock space for this system will not have the form of a tensor product of different Hilbert spaces associated with the masses and the string. Although this is quite natural from this perspective there are, however, physical questions that come to mind. For instance, if we think about this model as two masses connected by a physical spring (with ``internal degrees of freedom''), how does one recover the situation where the string just models an ideal spring? What is the origin of the $L^2(\mathbb{R})\otimes L^2(\mathbb{R})$ Hilbert space that one would use to describe this system? As a first step towards addressing these questions it is important to understand in detail why the Fock space does not factorize. We do this in the following.}

\subsection{Construction of the Fock space}

An accepted way to quantize linear systems relies on the construction of a Fock space. In the specific case of quantum field theory in curved spacetimes the relevant details can be found in \cite{Wald}. The starting point is the real vector space $\mathcal{S}$ of the classical (smooth) solutions to the field equations. By introducing a complex structure (among the many available) a complexified version of $\mathcal{S}$ is defined and endowed with a sesquilinear form obtained with the help of the symplectic structure. Taking a subspace of ``positive frequency'' solutions, the sesquilinear form defines a proper scalar product $\langle\cdot{},\!\cdot{}\rangle_+$. By completing the  complex vector space of positive frequency solutions w.r.t.\ this scalar product we obtain the one particle Hilbert space $\mathfrak{h}$. Finally, the Hilbert space of states of the quantum field theory is the Fock space $\mathcal{H}:=\mathcal{F}(\mathfrak{h})$.

In our case, the Hamiltonian description of the preceding section has produced a linear manifold domain of $L^2_\mu\times L^2_\mu$ given by $\mathcal{M}_2=\widehat{\mathcal{D}}\times \mathcal{D}$, where the classical Hamiltonian dynamics takes place, and a Hamiltonian vector field $X$ tangent to the closure of $\mathcal{M}_2$. Notice that we can pullback the canonical symplectic form in phase space
\begin{equation}\label{symplectic_2}
\Omega_{(Q,P)}((q_1,p_1),(q_2,p_2))=\langle q_1,p_2\rangle_\mu-\langle q_2,p_1\rangle_\mu
\end{equation} to $\mathcal{M}_2$ (where $(Q,P),(q_i,p_i)\in\mathcal{M}_2$).
We introduce now a complexification of this vector space  $\mathcal{M}_2^{\mathbb{C}}$ and use the symplectic form to define a scalar product. Vector addition is defined componentwise as the standard sum of real functions and multiplication by complex scalars is defined by introducing the complex structure:
\begin{equation}
\label{complexstr}
\mathbf{J}:\mathcal{M}_2\times \mathcal{M}_2\rightarrow \mathcal{M}_2\times \mathcal{M}_2:\big((q_1,p_1),(q_2,p_2)\big)\mapsto \big((-q_2,-p_2),(q_1,p_1)\big)
\end{equation}
and requiring that
\begin{equation}
(a+bi)\cdot V:=(a\mathbf{I}+b\mathbf{J})(V)
\end{equation}
for $a,b\in \mathbb{R}$ and $V\in \mathcal{M}_2\times \mathcal{M}_2$. In this case one can think of the elements in the complexified vector space as complex functions in $\mathcal{M}_2$ with the standard sum and multiplication by complex scalars. The complexified symplectic form is the straightforward extension by complex linearity of \eqref{symplectic_2}
\[
\Omega^{\mathbb{C}}_{(Q,P)}((q_1,p_1),(q_2,p_2)):=\langle q_1,p_2\rangle_\mu-\langle q_2,p_1\rangle_\mu\,.
\]
The integral curves of the Hamiltonian vector field $X$ can be identified with the space of solutions to the equations of motion. As $X$ is defined in terms of the elliptic operator $\Delta_\mu$, its eigenvalues $\lambda_n$ and normalized eigenfunctions $Y_n$ satisfying $\Delta_\mu Y_n=\lambda_nY_n$  (see appendix \ref{section_appendix_normal_modes}), play a relevant role in the following. It is important to point out here that these are not necessarily classical, regular solutions (for instance, they are not smooth with compact support). Furthermore, as the evolution is a symplectic transformation, the symplectic form can be pulled back to this space in the obvious way. In particular, the complexified solution defined by the (complex) Cauchy data $(Q,P):=(u(\cdot,0),\dot{u}(\cdot,0))\in\mathcal{M}_2^{\mathbb{C}}$ at $t=0$ can be written in the form
\begin{eqnarray*}
u(x,t)&=&\frac{1}{2}\sum_{n}\left(e^{it\sqrt{\widetilde{\omega}^2-\lambda_n}}\Big(Q_n-i\frac{P_n}{\sqrt{\widetilde{\omega}^2-\lambda_n}}\Big)
+e^{-it\sqrt{\widetilde{\omega}^2-\lambda_n}}\Big(Q_n+i\frac{P_n}{\sqrt{\widetilde{\omega}^2-\lambda_n}}\Big)\right)Y_n(x)
\end{eqnarray*}
with $Q_n=\langle Y_n,Q\rangle_\mu\in\mathbb{C}$ and $P_n=\langle Y_n,P\rangle_\mu\in\mathbb{C}$. Notice that, as mentioned in section \ref{classical_description}, our assumption that $\widetilde{\omega}>0$ implies that $\lambda_n<\widetilde{\omega}^2$. In order to select a subspace of positive frequency solutions we require that
\begin{equation}
iP+\sqrt{\widetilde{\omega}^2-\Delta_\mu}\,Q=0\,.
\label{condition_positive}
\end{equation}
This condition is equivalent to
\[
Q_n+i\frac{P_n}{\sqrt{\widetilde{\omega}^2-\lambda_n}}=0\,,\,\quad\forall n\,.
\]
It is now straightforward to see that, when equation \eqref{condition_positive} holds, we can indeed define a scalar product in this ``positive frequency part'' of the solution space as
\begin{equation}
\langle Q^{(1)},Q^{(2)}\rangle_+:=-i\Omega^{\mathbb{C}}((\overline{Q^{(1)}},\overline{P^{(1)}}),(Q^{(2)},P^{(2)}))=2\langle \overline{Q^{(1)}}, \sqrt{\widetilde{\omega}^2-\Delta_\mu}\,Q^{(2)}\rangle_\mu\,.
\label{scalar_product_solution_space}
\end{equation}
At this point, the only remaining step to finish the construction of $\mathfrak{h}$ is to Cauchy complete in the norm defined by the scalar product. In terms of the Fourier coefficients of $Q^{(1)}$ and $Q^{(2)}$ in the orthonormal basis that we are using the preceding scalar product becomes
\[
\langle Q^{(1)},Q^{(2)}\rangle_+=2\sum_n \sqrt{\widetilde{\omega}^2-\lambda_n}\,\,\langle\overline{ Y_n, Q^{(1)}}\rangle_\mu\langle Y_n,Q^{(2)}\rangle_\mu.
\]
Notice that the $L^2_\mu$-orthonormal basis $\{Y_n\}$ leads to an orthonormal basis of $\mathfrak{h}$ consisting of
\[
Z_n:=\frac{1}{\langle Y_n,Y_n\rangle_+^{1/2}}\,Y_n=\frac{1}{\sqrt{2}(\widetilde{\omega}^2-\lambda_n)^{1/4}}\,Y_n\,,
\]
hence, the one particle Hilbert space can be identified with
\[
\mathfrak{h}=\left\{\psi=\sum_n \psi_nZ_n\,\,:\, \psi_n\in \mathbb{C}\,,\,\sum_n|\psi_n|^2<\infty\right\}\,.
\]
Finally the Hilbert space $\mathcal{H}$ of our quantum field theory is given by the symmetric Fock space over $\mathfrak{h}$, $\mathcal{H}=\mathcal{F}(\mathfrak{h})$.

The standard procedure described in \cite{Wald} can be used to introduce creation and annihilation operators, quantum fields and the quantum Hamiltonian  $\hat{H}$ that generates the quantum dynamics of the system. In particular $\hat{H}$ is given by the lift to the Fock space of $\sqrt{\widetilde{\omega}^2-\Delta_\mu}$. Notice that the unitarity of the quantum evolution is guaranteed by the self-adjointness of the operator $\Delta_\mu$ in $\widehat{\mathcal{D}}$.

\subsection{Factorization of the Fock space}
In this section we will discuss the impossibility of factorizing $\mathcal{H}=\mathcal{H}_{\mathrm{masses}}\otimes\mathcal{H}_{\mathrm{string}}$ in a \textit{natural} way, i.e.\ with ``factors'' associated,  respectively, with the point masses and the string. For the sake of the argument let us see what would happen if the Hilbert space for our model were the Fock space over $L^2_\mu$. As discussed in appendix \ref{subsection_L^2_mu}, the space $L^2_\mu$ is isomorphic to $\mathbb{R}\oplus L^2(0,1)\oplus\mathbb{R}$ where each $\mathbb{R}$ is associated with a point mass. Using the properties of the Fock construction it is straightforward to see that $\mathcal{F}(L^2_\mu)=\mathcal{F}(\mathbb{R})\otimes\mathcal{F}(L^2(0,1))\otimes\mathcal{F}(\mathbb{R})$ and, hence, if this were the case, we would have a Hilbert space corresponding to each of the point masses contributing a factor to the Fock space.

However we have to consider $\mathfrak{h}$, instead of $L^2_\mu$, and we are prescribed to use the scalar product $\langle\cdot{},\!\cdot{}\rangle_+$ involving the square root factor $\sqrt{\widetilde{\omega}^2-\Delta_\mu}$. This changes crucially the outcome and prevents us, in particular, to write $\mathfrak{h}=\mathbb{C}\oplus\mathfrak{h}_{\mathrm{string}}\oplus\mathbb{C}$ with the $\mathbb{C}$ subspaces associated with the boundaries as in the previous example. Indeed, if such a decomposition were available, the function
\begin{equation}
F:[0,1]\rightarrow \mathbb{R}:x\mapsto F(x)=\left\{\begin{array}{lcl}1&\ &x=0\\0&&x\in(0,1]\end{array}\right.
\end{equation}
considered in the appendix (see equation \eqref{f} and the discussion after it) would have to be in $\mathfrak{h}$. However, it is straightforward to see that, in the limit $n\rightarrow\infty$, the coefficients $\langle F,Z_n\rangle_+$ satisfy
\[
\langle F,Z_n\rangle_+=\frac{2\sqrt{\alpha_0}}{\sqrt{\mu_0 \pi n}}+O\left(n^{-5/2}\right)\,,
\]
so they are not square summable and, hence, $F\not\in\mathfrak{h}$.

The impossibility of achieving the previous decomposition immediately implies the non-factorizability of the Fock space built from $\mathfrak{h}$. Of course, this does not exclude the possibility of finding such a factorization in other ways, for example for every decomposition of the type $\mathfrak{h}=\mathfrak{h}_1\oplus \mathfrak{h}_2$ we would have
\[
\mathcal{H}=\mathcal{F}(\mathfrak{h})=\mathcal{F}(\mathfrak{h}_1\oplus \mathfrak{h}_2)=\mathcal{F}(\mathfrak{h}_1)\otimes\mathcal{F}(\mathfrak{h}_2)\,,
\]
however, they are definitely not obvious from the present perspective and, in particular, there is no reason \textit{a priori} to associate the factors to the point masses.

\section{Conclusions}\label{conclusions}

We have studied the Hamiltonian formulation and Fock quantization for a 1+1 dimensional model containing both fields and point masses. The combined description of different types of physical objects poses some interesting and non-trivial questions related to the proper characterization of the physical degrees of freedom both at the classical and quantum levels. Some natural questions for these systems are: Is it possible to talk about \textit{independent} degrees of freedom associated with the masses and the fields? or, is it possible to split the quantum Hilbert space of the system in the form $\mathcal{H}=\mathcal{H}_{\mathrm{masses}}\otimes\mathcal{H}_{\mathrm{field}}$? One of the main results of the paper is the proof that the Fock quantization of such compound model leads to a Fock space that cannot be written in a natural way as the tensor product of Hilbert spaces associated with the masses and the field respectively. We want to emphasize that this result comes about after a careful discussion of the spaces of solutions to the equations of motion and the details of the construction of the Fock space, in particular, the scalar product appearing in the complexified solution space. This lack of factorization {\color{black} must be understood} because one of the assumptions used in the description of compound quantum systems is that their state spaces are tensor products of Hilbert spaces associated with each subsystem. This statement is singled out by some authors as the zeroth postulate of Quantum Mechanics \cite{Zurek1,Zurek2}. Of course, in the present case our construction does not violate this postulate but, rather, shows that identifying subsystems in a proper way is not straightforward, even at the classical level.

From a technical point of view the method that we have employed relies on the introduction of Hilbert spaces endowed with scalar products defined with the help of modified measures describing the coupling of the fields with lower dimensional objects (point masses in the present case). This is useful both at the classical and quantum levels because the identification of the relevant functional spaces is simplified in a significant way. One of the key elements of our approach relies on the ideas developed by Evans \cite{Evans} to deal with modified Sturm-Liouville problems of the type considered here. It is especially important to work with elliptic, self-adjoint operators (generalized Laplacians) with spectrum and associated eigenfunctions that can be mapped to the ones appearing in the resolution of the non-standard eigenvalue problem that determines the normal modes of the system. These generalized Laplacians play a fundamental role in the Hamiltonian formulation of the model and its subsequent Fock quantization.

The Hamiltonian formulation that we have developed starts from a Lagrangian written in terms of the natural objects: measures, the scalar product defined with their help and the associated Radon-Nikodym derivatives. A striking feature of the approach that we follow is that although the constants appearing in the scalar product are functionally dependent on the physical parameters they are, generically, non-trivial functions of them. It is also important to emphasize the fact that the appropriate Lagrangians are non-trivial when written in terms of these objects because of the necessity to have appropriate self-adjoint operators (in particular it is very important that the normal modes of the system can be interpreted as eigenfunctions of the self-adjoint operator $\Delta_\mu$ with eigenvalues given by the normal frequencies).

We have described the construction of the Fock space by relying of the methods customarily used in the context of quantum field theory in curved spacetimes (in particular those described by Wald in \cite{Wald}). An advantage of combining those methods with the Hamiltonian formulation that we are using here is the possibility of having a precise and explicit characterization of the solution space for the equations describing the dynamics. Basically the only difference with the present case is due to the fact that the relevant elliptic self-adjoint operator is not a Laplace-Beltrami operator but one defined with the help of a measure with singular contributions at the boundaries of the region where the fields are defined.

We have illustrated the ideas developed in the paper with a simple 1+1 dimensional model but they can be extended without difficulty to other more complicated systems, in particular, our approach can be used to deal with higher-dimensional models and is flexible enough to allow the coupling of different types of low-dimensional objects not restricted to point masses. The model that we have studied provides an interesting way to interpolate between different types of boundary conditions (by, for instance, considering the limits of small or large masses at the boundary). It is important to highlight here that, the present methods can be of use not only for linear systems but also for other non-linear ones that are obtained by consistently adding interactions to them (for instance, gauge theories). By working with fields defined in unbounded space-time regions these techniques can also be adapted to the study of quantum dissipative systems (in the spirit of the Caldeira-Legget or Rubin models) and decoherence.

A final comment concerns the interpretation of the current system in terms of so-called particle detectors, which provide a clear-cut definition of a particle in general situations, such as curved spacetimes and non-inertial frames. In this approach, the particle detector is a system with additional degrees of freedom that interacts with the field according to some specific interaction Hamiltonian, as in e.g. \cite{Unruh:1976db, DeWitt, UnruhWald}. The effect of the interaction is read in the evolution of the states of the Hilbert space, which is a tensor product, and in which excitations and de-excitations in the sector of the detector according to its interaction with the field are interpreted as particles. For an in-depth review of the state of the art see \cite{Hu:2012jr}. As we have shown this type of factorization does not take place for such natural models as the one considered here so it is very important to understand and characterize the physical systems for which this is possible and the ensuing implications for the measurement problem in quantum mechanics.

%%%%%%%%%%%%

\appendix

\section{Hilbert spaces, modified measures and Radon-Nikodym derivatives}\label{Hilbert_spaces}
\renewcommand{\thesubsection}{\Alph{section}.\Roman{subsection}}
\subsection{The Hilbert space \texorpdfstring{$L^2_\mu$}{L2}}\label{subsection_L^2_mu}

The implementation of the GNH algorithm requires the introduction of appropriate Hilbert spaces. After realizing that a scalar product of the type \eqref{scalar_product} plays a natural role in the eigenvalue problems that crop up in the study of the system that we are considering here, it is natural to construct Hilbert spaces endowed with it. To see that this is not a completely trivial task, it suffices to realize that \eqref{scalar_product} does not make sense for elements of $L^2(0,1)$ as they do not have well defined boundary values at $x=0,1$ (because points have zero measure with respect to the standard Lebesgue measure). A possible way out could be to consider other types of Hilbert spaces, for instance the Sobolev space $H^1(0,1)$. The extra regularity of the elements in Sobolev spaces makes it possible to define their boundary values by using the so called \textit{trace operator} $\gamma$. The elements of the $H^1(I)$ Sobolev space defined on an interval $I\subset \mathbb{R}$ have representatives given by continuous functions on its closure $\overline{I}$ \cite{Brezis}. Their values at the boundary (or their limits for that matter) are the traces that we denote as  $\gamma_0$, $\gamma_1$ because the boundary of an interval is disconnected.

In the present context it is better to avoid demanding so much regularity so we consider, to begin with, spaces with elements satisfying minimal regularity requirements. The simplest example of such a space would be $\mathbb{L}^2[0,1]:=\mathbb{R}\times L^2(0,1)\times \mathbb{R}$ consisting of elements that we will denote as $\vec{v}:=(v_0,v,v_1)$. This is a Hilbert space with the scalar product
\begin{equation}
\langle \vec{v},\,\vec{w} \rangle_{\mathbb{L}}:=\alpha_0 v_0w_0+\alpha_1v_1w_1+\int_{[0,1]} v\cdot w\, \mathrm{d}\mu_L\,
\label{scalar_product_2}
\end{equation}
provided that $\alpha_j>0$. Notice that the completeness of $\mathbb{L}^2[0,1]$ is a direct consequence of that of $L^2(0,1)$ and $\mathbb{R}$.

A useful alternative way to describe this space makes use of a suitable measure $\mu$. Let us consider the Borel $\sigma$-algebra $\mathfrak{B}([0,1])$ and the measure
 $\mu=\alpha_0 \delta_0+\mu_L|_{(0,1)}+\alpha_1\delta_1$, where the measure $\delta_j$ is defined for any $A\in\mathfrak{B}([0,1])$ by
\begin{equation}
\delta_j(A)=\left\{\begin{array}{ll}1&\textrm{if}\ j\in A\\0&\textrm{if}\ j\notin A\end{array}\right.
\label{Diracdelta}
\end{equation}
Let us consider now the real Hilbert space of square integrable functions
\[
L^2_\mu[0,1]:=\left\{f:[0,1]\rightarrow \overline{\mathbb{R}}\ : \ f\, \mathrm{is} \, \mu\textrm{-measurable}\,,\,\int_{[0,1]}f^2\, \mathrm{d}\mu<+\infty\right\}\,,
\]
(as usual, functions equal $\mu$-a.e.\ are identified) endowed with the scalar product
\begin{equation}
\langle u,v \rangle_\mu=\int_{[0,1]}u\cdot v\, \mathrm{d}\mu\,.
\label{scalar_product_3}
\end{equation}
It is straightforward to see that the map

\[\begin{array}{cccc}
\Phi:&L^2_\mu[0,1] & \longrightarrow & \mathbb{L}^2[0,1]\\
     &     f       &    \longmapsto  & (f(0),\left.f\right|_{(0,1)},f(1))
\end{array}\]
is a Hilbert space isomorphism, hence, in the following we will view elements of these Hilbert spaces as square integrable functions (w.r.t.\ the measure $\mu$) defined on the closed interval $[0,1]$ or as elements of  $\mathbb{R}\times L^2(0,1)\times \mathbb{R}$ with the scalar products defined above. Whenever no confusion may arise we will use the shorthand $L^2_\mu$ to denote $L^2_\mu[0,1]$.

\subsection{Absolutely continuous functions and Radon-Nikodym derivatives}\label{appendixB2}

A useful way to write the Lagrangians that we use in the paper makes use of the Radon-Nikodym (RN) derivatives of appropriately defined functions. They also play a central role in the description of the relevant elliptic operators that appear in the Hamiltonian formulation and the construction of the Fock space. Here we briefly review the central concepts and give a list of properties that are used throughout the paper.

Given two measures $\mu$ and $\nu$ over $\mathfrak{B}([0,1])$, we say that $\nu$ is $\mu$-absolutely continuous ($\mu$-a.c.\ usually denoted as $\nu\ll\mu$) if whenever $\mu(A)=0$ for some $A\in\mathfrak{B}([0,1])$, then $\nu(A)=0$. The RN theorem states that over finite measure spaces, such definition is equivalent to ``weak $\mu$-differenciability'', which means that there exists a $\mu$-measurable function $f\in L^1(\mu)$ such that
\[
\nu_f(A)=\int_A f\, \mathrm{d}\mu\,.
\]
The function $f$ (usually denoted by $f=\frac{\mathrm{d}\nu}{\mathrm{d}\mu}$) is known as the RN derivative and it is unique in the sense that any other function that satisfies the preceding properties is equal to $f$ $\mu$-a.e.

If we want to $\mu$-differentiate a function $F:[0,1]\rightarrow\mathbb{R}$ w.r.t.\ $\mu=\alpha_0\delta_0+\left.\mu_L\right|_{(0,1)}+\alpha_1\delta_1$ we have to associate to $F$ a $\mu$-a.c.\ Lebesgue-Stieltjes  measure $\nu_F$. To this end we define
\[
\nu_F=\Big(F(0+)-F(0)\Big)\delta_0+F'\!\left.\mu_L\right|_{(0,1)}+\Big(F(1)-F(1-)\Big)\delta_1
\]
where $\left.F\right|_{(0,1)}$ is a.c.\ in the usual calculus sense (so that $F'\!\left.\mu_L\right|_{(0,1)}$ is $\mu_L$-a.c.).   Under these conditions $F$ is differentiable $\mu_L$-a.e.\ and has well defined limits $F(0+)$ and $F(1-)$, not necessarily equal to its values at the boundary $F(0)$ and $F(1)$. In fact, considering that the a.c.\ functions over $(0,1)$ can be seen as elements of the Sobolev space $H^1(0,1)$ (functions of $L^2(0,1)$ with distributional derivative in $L^2(0,1)$), and also that no restriction over the boundary $\{0,1\}$ is placed by the condition of being $\mu$-a.c.\ continuous, we conclude that the vector space of $\mu$-a.c.\ functions $H^1_\mu[0,1]$ is isomorphic to $\mathbb{R}\times H^1(0,1)\times\mathbb{R}$ (compare with $L^2_\mu[0,1]\cong\mathbb{R}\times L^2(0,1)\times\mathbb{R}$). An element $F\in H^1_\mu[0,1]$ has boundary values $F(0)$, $F(1)$ and traces $\gamma_0(F)=F(0+)$, $\gamma_1(F)=F(1-)$ which are generically different from $F(0)$ and $F(1)$. This is related to the fact that, in general, $\alpha_j\neq\mu_j$. In our model the boundary values of functions (such as $F(0)$, $F(1)$) play a secondary role \textit{from the physical point} of view as the positions of the point masses will be defined by the traces, but they play a central role \textit{from the mathematical point of view} because they are crucial to prove that the fundamental Laplace operator of the model is self-adjoint.

Finally we define the RN derivative of $F\in H^1_\mu[0,1]$ as $\frac{\mathrm{d}F}{\mathrm{d}\mu}:=\frac{\mathrm{d}\nu_F}{\mathrm{d}\mu}$. Notice that if $\frac{\mathrm{d}F}{\mathrm{d}\mu}$ is in $H^1_\mu[0,1]$, we can apply again the RN theorem to obtain another derivative that we denote $\frac{\mathrm{d}^2F}{\mathrm{d}\mu}:=\frac{\mathrm{d}}{\mathrm{d}\mu}\left(\frac{\mathrm{d}F}{\mathrm{d}\mu}\right)$. The explicit form of the RN derivative of $F\in H^1_\mu[0,1]$ can be immediately obtained from the previous result by comparing the expressions of $\nu_F$ and $\mu$
\begin{itemize}
    \item $\dfrac{\mathrm{d}F}{\mathrm{d}\mu}(j)=\dfrac{(-1)^j}{\alpha_j}\big(\gamma_j(F)-F(j)\big)$.
    \item $\dfrac{\mathrm{d}F}{\mathrm{d}\mu}(x)=F'(x)$, $\mu_L\textrm{-a.e.}$ in $(0,1)$ which is just the RN derivative of $F$ w.r.t.\ $\mu_L$.
\end{itemize}
It is important to notice that the standard Leibniz rule does not hold, in general, for RN derivatives though a modified rule exists. Indeed, by defining $K:[0,1]\rightarrow \mathbb{R}$ with $\left.K\right|_{(0,1)}=0$ and $K(j)=(-1)^j\alpha_j$ for $j\in\{0,1\}$ we have:
        \begin{equation}
          \frac{\mathrm{d}(FG)}{\mathrm{d}\mu}(x)=\frac{\mathrm{d}F}{\mathrm{d}\mu}(x)G(x)+F(x)\frac{\mathrm{d}G}{\mathrm{d}\mu}(x)+K(x)\frac{\mathrm{d}F}{\mathrm{d}\mu}(x)\frac{\mathrm{d}G}{\mathrm{d}\mu}(x)\,,\quad \mu\textrm{-a.e.}\,\,\mathrm{in}\,\,[0,1]\,.\label{intbyparts}
        \end{equation}
This expression will be very useful whenever we have to perform integrations by parts of RN derivatives.

\subsection{The fundamental Laplace operator}\label{section_appendix_laplacian}
Next we will introduce a generalized Laplace operator, defined in terms of the RN derivative, that plays a central role in the paper. Let us write
\begin{eqnarray}
&&\mathcal{D}:=\left\{u\in L^2_\mu[0,1]:\,\exists\dfrac{\mathrm{d}u}{\mathrm{d}\mu}(x) \ \forall x\in[0,1]\,, \dfrac{\mathrm{d}u}{\mathrm{d}\mu}{\mathrm{\ is}\,\mu\textrm{-a.c.}}\,, \dfrac{\mathrm{d}^2u}{\mathrm{d}\mu^2}\in L^2_\mu[0,1]\right\}\,,\label{domain_1}\\
&&\widehat{\mathcal{D}}:=\left\{u\in \mathcal{D}:(-1)^j\dfrac{\mathrm{d}u}{\mathrm{d}\mu}(j)-A(j)u(j)=0\right\}\,,\label{domain_2}\\[1ex]
&&\Delta_\mu:\mathcal{D}\subset L^2_\mu[0,1]\rightarrow L^2_\mu[0,1]:\ u\mapsto (1+C)\dfrac{\mathrm{d}^2u}{\mathrm{d}\mu^2}\,,
\label{Laplacian_gen}
\end{eqnarray}
where $C:[0,1]\rightarrow \mathbb{R}$ and $A(j)\in\mathbb{R}$ for $j\in\{0,1\}$ are to be determined together with $\alpha_j$.  One condition that we have to impose is that we should recover the equations of motion, in particular for the \textit{lateral limits} at the boundaries. Fulfilling this requirement does not completely fix these constants, but puts the eigenfunctions of $\Delta_\mu$ in correspondence with the ones of the original problem. Hence, if we make $\Delta_\mu$ self-adjoint, then the set of eigenfunctions will form a complete set of orthogonal functions w.r.t.\ the scalar product $\langle\cdot{},\!\cdot{}\rangle_\mu$.

So let us compare equations \eqref{ee1}-\eqref{ee4} with the eigenvalue problem {\color{black}$\Delta_\mu Y=\lambda Y$} (defined in a subspace of $L^2_\mu$). First, \eqref{ee1} requires that $\left.C\right|_{(0,1)}=0$ because then $\Delta_\mu$ is the usual Laplacian (second derivative) when restricted to the interval $(0,1)$. Taking $C(j)=A(j)\alpha_j$ makes $\Delta_\mu$ symmetric (i.e.\ $\langle \Delta_\mu u,w\rangle_\mu-\langle u,\Delta_\mu w\rangle_\mu=0$) provided we restrict its domain to $\widehat{\mathcal{D}}$. In fact using \eqref{intbyparts} we have:
\begin{align*}
\langle \Delta_\mu u,w\rangle_\mu&=-\left\langle \frac{\mathrm{d}u}{\mathrm{d}\mu},\frac{\mathrm{d}w}{\mathrm{d}\mu}\right\rangle_{\hspace*{-1ex}\mu\hspace*{0.1ex}}
-\sum_{j\in\{0,1\}}(-1)^j\frac{\mathrm{d}u}{\mathrm{d}\mu} w-\sum_{j\in\{0,1\}}\alpha_j\frac{\mathrm{d}^2u}{\mathrm{d}\mu^2}\left((-1)^j\alpha_j \frac{\mathrm{d}w}{\mathrm{d}\mu}-C(j) w\right)\\
&=-\left\langle \frac{\mathrm{d}u}{\mathrm{d}\mu},\frac{\mathrm{d}w}{\mathrm{d}\mu}\right\rangle_{\hspace*{-1ex}\mu\hspace*{0.1ex}}
+\sum_{j\in\{0,1\}}A(j)u(j)w(j)\,.
\end{align*}
where we have used that the boundary conditions $(-1)^j\frac{\mathrm{d}u}{\mathrm{d}\mu}(j)-A(j)u(j)=0$ and our particular choice of $C(j)$, leading to a symmetric expression in $u$ and $w$.

It is important to notice that ussing the expression given for the RN derivative at the boundary, we can show that the Robin-like boundary conditions $(-1)^j\frac{\mathrm{d}u}{\mathrm{d}\mu}(j)-A(j)u(j)=0$ are equivalent to
\begin{equation}
    \gamma_j(u)=(1+\alpha_jA(j))u(j)\,.\label{eq robin-like equivalent}
\end{equation}
Now taking $A(j)=\frac{\mu_j\Delta\widetilde{\omega}^2_j}{1-\alpha_j\mu_j\Delta\widetilde{\omega}^2_j}$ and using the previous equation for $\frac{\mathrm{d}F}{\mathrm{d}\mu}$ and for $\gamma_j(F)$, we obtain that $\Delta_\mu Y=\lambda Y$ on $(0,1)$ is equivalent to:
\begin{eqnarray}
    &&Y''=\lambda Y\,,\\
    &&\gamma_j(Y')=(-1)^j\Big(\alpha_j(1-\alpha_j\mu_j\Delta\widetilde{\omega}^2_j)^2\lambda+\mu_j\Delta\widetilde{\omega}^2_j\Big)\gamma_j(Y)\,.
     \label{frontera_1}
\end{eqnarray}
If we finally require $\alpha_j(1-\alpha_j\mu_j\Delta\widetilde{\omega}^2_j)^2=\mu_j$, equations \eqref{frontera_1} become equations \eqref{ee3}-\eqref{ee4}, recovering thus the original problem. Clearly from this condition and the fact that $\mu_j>0$, we see that all the possible solutions for $\alpha_j$ are positive, so they can be used to define scalar products of the form \eqref{scalar_product_2},\eqref{scalar_product_3}. A very simple case corresponds to $\Delta\widetilde{\omega}_j=0$ (some sort of resonance that cancels out the effect of the springs and the string) for which $\alpha_j=\mu_j$ and $A(j)=0$. It is important to realize that, as we have already mentioned before, the positions of the point masses are given by the traces {\color{black}$\gamma_j(Y)$ (which can be identified with the $X(j)$ of section \ref{classical_description}), however, the values $Y(j)$ that appear in the Robin-like boundary conditions \eqref{eq robin-like equivalent} \textit{do not have a direct physical interpretation} as far as we know.}

At this point, we have only managed to obtain a symmetric operator $\Delta_\mu$ whose eigenfunctions are in correspondence with the ones of the original problem, but in fact, it can be shown that the operator $\Delta_\mu$ is self-adjoint in $\widehat{\mathcal{D}}$ by using the method described in \cite{Evans}. As mentioned before, this implies that the set of its eigenfunctions can be turned into a Hilbert basis of the Hilbert space $\widehat{\mathcal{D}}$ showing in particular, that it is complete. A remark is in order here, if $u,v\in\widehat{\mathcal{D}}$ we have, using equation \eqref{eq robin-like equivalent}, that:
\begin{equation}\label{eq_equivalence_scalar}
\langle u,v \rangle_\mu=\sum_{j\in\{0,1\}}\alpha_j u(j)v(j)+\langle u,v\rangle=\sum_{j\in\{0,1\}}\mu_j \gamma_j(u)\gamma_j(v)+\langle u,v\rangle=\langle\!\langle u,v\rangle\!\rangle\,.
\end{equation}
This establishes a connection between the results in the present section with the argument leading to \eqref{scalar_product} (notice that the boundary values of the eigenfunctions appearing there can --and should-- be interpreted as right or left limits or, in our notation as $\gamma_j(u),\gamma_j(v)$).

\subsection{Normal modes and Fourier coefficients}\label{section_appendix_normal_modes}

An orthonormal (Hilbert) basis of $L^2_\mu[0,1]$ can be constructed by using the eigenfunctions of the self-adjoint operator $\Delta_\mu$ defined in \eqref{Laplacian_gen} in the domain $\widehat{\mathcal{D}}$  given in \eqref{domain_2}. These eigenfunctions, in turn, can be computed from the solutions to the eigenvalue problem considered in \eqref{ee1}-\eqref{ee4} by defining their values at $x\in\{0,1\}$ in such a way that the Robin-like boundary conditions, or equivalently equation \eqref{eq robin-like equivalent}, are satisfied. The elements of the Hilbert basis will, hence, be of the form $\{Y_n\}$ (in one to one correspondence with the eigenfunctions $\{X_n\}$ introduced on section \ref{classical_description}) with $\left.Y_n\right|_{(0,1)}=X_n/g_n$ (in particular $\gamma_j(Y_n)=X_n(j)/g_n$) and $Y_n(j)=(1-\alpha_j\mu_j\Delta\widetilde{\omega}^2_j)X_n(j)/g_n$. The factor $g_n$ guarantees that $Y_n$ is normalized w.r.t.\ the scalar product $\langle \cdot ,\!\cdot \rangle_\mu$. It can be explicitly computed to be
\[
g_n^2=\frac{1}{2}\left(\mu_0\Delta\widetilde{\omega}^2_0+(1+\mu_0)\omega^2_n
+\mu_0^2(\omega^2_n-\Delta\widetilde{\omega}^2_0)^2
+\mu_1^2(\omega^2_n+\Delta\widetilde{\omega}^2_1)\frac{\omega^2_n
+\mu_0^2(\omega^2_n-\Delta\widetilde{\omega}^2_0)^2}{\omega^2_n+\mu_1^2(\omega^2_n-\Delta\widetilde{\omega}^2_1)^2}\right)\,.
\]
The asymptotic behavior of $1/g_n$ when $n\rightarrow\infty$ can be obtained by using equation \eqref{asympt}:
\[
g_n^{-1}=\frac{\sqrt{2}}{\mu_0^2}\frac{1}{\pi^2n^2}+O\left(n^{-4}\right)\,.
\]

An element $F\in L^2_\mu[0,1]$ can be expanded as usual as $F=\sum \langle Y_n,F\rangle_\mu Y_n$. It is interesting to notice at this point that functions $f:[0,1]\rightarrow \mathbb{R}$ supported at the boundary $\{0,1\}$ are non trivial elements of $L^2_\mu[0,1]$ and hence can be expanded in the basis introduced above. In the particular case of the function $F\in L^2_\mu[0,1]$ given by
\begin{equation}
F:[0,1]\rightarrow \mathbb{R}:x\mapsto F(x)=\left\{\begin{array}{lcl}1&\ &x=0\\0&&x\in(0,1]\end{array}\right.
\label{f}
\end{equation}
we have $\langle Y_n,F\rangle_\mu=\alpha_0Y_n(0)=\alpha_0(1-\alpha_0\mu_0\Delta\widetilde{\omega}^2_0)X_n(0)/g_n=\sqrt{\mu_0\alpha_0}X_n(0)/g_n$. In the discussion of the Fock quantization of the model we need the asymptotic behavior of these coefficients for large values of $n$. This can be easily obtained by using the fact that in this limit only the eigenvectors associated with negative eigenvalues matter. The asymptotic behavior of $g_n^{-1}$ and the one of $\omega_n$ given in equation \eqref{asympt} lead to
\begin{align*}
   \langle Y_n,F\rangle_\mu&=\sqrt{\mu_0\alpha_0}\,\,\frac{X_n(0)}{g_n}=\sqrt{\mu_0\alpha_0}\,\,\frac{\omega_n}{g_n}=\sqrt{2\mu_0\alpha_0}\,\,\frac{1}{\pi n}+O\left(n^{-2}\right)\,.
\end{align*}

%%%%%%%%%%%%%%%%%%%%%%%%%%%%%%%%%%%%%%%%%%%%%%%%%%%%%%%%%%%%%%%%%%%%%%%
%
% ACKNOWLEDGMENTS
%
\section*{Acknowledgments}

We would like to thank J. Louko for some enlightening discussions. This work has been supported by the Spanish MINECO research grants FIS2012-34379, FIS2014-57387-C3-3-P and the  Consolider-Ingenio 2010 Program CPAN (CSD2007-00042). B. Ju\'arez-Aubry is supported by CONACYT, M\'exico REF 216072/311506 with additional support from Sistema Estatal de Becas, Veracruz, M\'exico. Juan Margalef-Bentabol is supported by a ``la Caixa'' fellowship.

%%%%%%%%%%%%

\end{document}